\documentclass[12pt]{article}
\usepackage[pdftex]{color}
\usepackage{cancel} %
\usepackage{amsmath,amsthm,amssymb}
\usepackage{wasysym}
\usepackage{graphicx}
\usepackage{float}
\usepackage{color}%
\usepackage[onehalfspacing]{setspace}
\usepackage{lipsum}
\usepackage[lmargin=2.5cm,rmargin=2.5cm,tmargin=3cm,bmargin=3cm]{geometry}
\usepackage{multirow}
\usepackage{hyperref}
\hypersetup{hidelinks}
\usepackage{cleveref}
\usepackage{accents}
\usepackage[ruled,vlined,linesnumbered]{algorithm2e}
\usepackage{cite}
\usepackage{lineno}
\usepackage{scalerel}
\usepackage{stackengine}
\usepackage{nomencl}
\makenomenclature
\stackMath
\definecolor{color}{rgb}{0.8500, 0.3250, 0.0980} 

\let\oldequation\equation
\let\oldendequation\endequation
\renewenvironment{equation}
{\linenomathNonumbers\oldequation}
{\oldendequation\endlinenomath}

\begin{document}
\begin{center}
{\rm \bf \Large{Influence of twist direction and large deformation on soft material torsional contact}}
\end{center}

\begin{center}
{\bf Yucai Hu}$^{\text{a}}$, {\bf Pengfei Li}$^{\text{a}}$, {\bf Michele Ciavarella}$^{\text{b}}$, {\bf Yue Wu}$^{\text{c}}$\footnote{Corresponding author: wuy@sari.ac.cn}, {\bf Yang Xu}$^{\text{ad}}$\footnote{Corresponding author: yang.xu@hfut.edu.cn}
\end{center}
\begin{flushleft}
$^{\text{a}}$School of Mechanical Engineering, Hefei University of Technology, Hefei, 230009, China \\
$^{\text{b}}$Center of Excellence in Computational Mechanics, Politecnico di BARI, Viale Japigia 182, 70126 Bari, Italy \\
$^{\text{c}}$Shanghai Advanced Research Institute, Chinese Academy of Sciences, Shanghai 201200, China \\
$^{\text{d}}$Anhui Province Key Laboratory of Digital Design and Manufacturing, Hefei, 230009, China \\
\end{flushleft}

\begin{abstract}
Shear-induced contact area reduction is widely observed in soft contacts, yet recent torsional experiments have revealed a more complex non-monotonic evolution in which the contact area first increases and then decreases with twist angle. The mechanism responsible for this initial area increase and the role of large deformation in the overall area evolution remain unclear. In this study, we experimentally investigate the torsional contact response of soft Polydimethylsiloxane (PDMS) spheres by combining forward-backward twist tests with a systematic variation of the curing-agent-to-base ratio to tune material softness and deformation level. The loading-unloading tests show that the torsional interface is strongly irreversible: during unloading, the contact area follows a decrease-increase-decrease path rather than retracing the loading branch, and repeatable petal-like edges appear, indicating a wrinkle-induced surface instability. By decreasing the mixing ratio, we find that larger deformation strengthens the area-reduction contribution and eventually suppresses the initial area increase, leading to a monotonic area decrease during loading for sufficiently soft PDMS. Softer PDMS also exhibits lower shear strength, weaker torque oscillations, and improved repeatability. The results provide experimental evidence that large deformation can drive shear-induced contact area reduction, while the origin of the initial area increase remains unresolved. These findings narrow the possible mechanisms (e.g., triboelectrification) responsible for the initial area increase and provide a stringent benchmark for frictional contact models of soft interfaces.

\end{abstract}


\section{Introduction}

The variation of contact area at a frictional interface is central to many tribological phenomena in both nature and industry, including metainterfaces \cite{aymard2024designing,zeka2026normal}, earthquakes \cite{scholz1976asperity}, seal crawling in hydraulic systems \cite{nikas2003elastohydrodynamics}, and rubber pad walking in rail fastening systems \cite{zeng2020experimental}. In soft contacts, shear-induced contact area reduction has been extensively observed as a fundamental response of an elastomer sliding against a rigid counter-surface \cite{savkoor1977effect, waters2010mode, sahli2018evolution, mergel2018continuum,wei2024does,wei2025characterizing}. For a smooth spherical contact, the contact area typically shrinks from a circle to an ellipse as the tangential load increases, until the entire interface enters sliding. A similar shear-induced area reduction has also been reported for rough contacts \cite{sahli2018evolution}. However, Zhang et al. \cite{zhang2024non} recently reported a torsional contact experiment between a PDMS sphere and a flat glass substrate that departs significantly from this established picture. In that study, the contact area evolves non-monotonically with the twist angle, first increasing and then decreasing until the entire interface is in sliding. A similar non-monotonic trend was also observed in unidirectional shear and later confirmed independently by Wei et al. \cite{wei2024does} (see Fig. 5 in \cite{wei2024does}). Moreover, Zhang et al. \cite{zhang2024non} showed that the initial area increase is insensitive to twisting speed, normal force, and viscoelasticity. Related observations have also emerged in other systems. Ahamad et al. \cite{ahmad2024effect} found in finger-pad sliding that the contact area may either decrease or increase depending on the sliding direction, which they attributed to non-uniform local curvature of the interface. 

Ever since the seminal work of Savkoor and Briggs \cite{savkoor1977effect}, the coupling between friction and adhesion has been regarded as a primary mechanism driving shear-induced contact area reduction \cite{waters2010mode, papangelo2019mixed, mcmeeking2020interaction, xu2022asperity}. Existing linear-elastic-fracture-mechanics-based adhesive friction models reproduce the experimentally observed decrease in contact area with tangential loading reasonably well. However, adhesion may not be the only mechanism involved. Lengiewicz et al. \cite{lengiewicz2020finite} and Mergel et al. \cite{mergel2021contact} showed that large deformation contact models can also produce shear-induced area reduction even in the absence of adhesion. More recently, Ceglie et al. \cite{ceglie2025contact} developed a wavy-contact model under large deformation and found that, at high amplitude-to-wavelength ratios, a soft waviness flattened by a rigid flat can exhibit a non-monotonic area evolution similar to that reported by Zhang et al. \cite{zhang2024non}. These results suggest that large deformation may play an important role, while adhesion may amplify the effect rather than fully govern it. In parallel, Xu et al. \cite{xu2025analytical} developed a new analytical framework for predicting the contact area reduction, indicating that the corresponding driver is Poynting's effect. The aforementioned studies show that the decrease in contact area under shear can arise from several coupled mechanisms. However, for torsional soft contact, the origin of the initial increase in contact area remains unresolved, and the specific role of large deformation is still unclear.

The present work advances beyond Zhang et al. \cite{zhang2024non} in three key aspects. First, while Zhang et al. \cite{zhang2024non} investigated only the forward twist stage, we perform complete forward-backward torsional cycles to examine the reversibility of the interface. Hereafter, the loading and unloading stages are used to refer to the forward and backward twists, respectively. Second, we demonstrate that the unloading path is strongly irreversible: the contact area follows a decrease-increase-decrease path rather than retracing the loading branch, and repeatable petal-like edges emerge, indicating a wrinkle-induced surface instability intrinsic to the soft elastomer. Third, Zhang et al. \cite{zhang2024non} used only a single PDMS mixing ratio ($1:10$), whereas we systematically vary the curing-agent-to-base ratio across four levels ($1:5$, $1:10$, $1:15$, $1:20$) to tune the extent of large deformation and isolate its role in the contact-area evolution. 

The remainder of the paper is organized as follows. Section \ref{sec:Test} introduces the torsional contact test rig and the mechanical properties of PDMS with different mixing ratios obtained from uniaxial tensile tests. Sections \ref{sec:Oscillating_contact} and \ref{sec:Large_deformation} present the effects of loading direction and large deformation on the mechanical response of the torsional interface, respectively, followed by a discussion in Section \ref{sec:Discussion} and a conclusion in Section \ref{sec:Conclusion}.

\section{Torsional contact test}\label{sec:Test}
\begin{figure}[H]
  \centering
  \includegraphics[width=15cm]{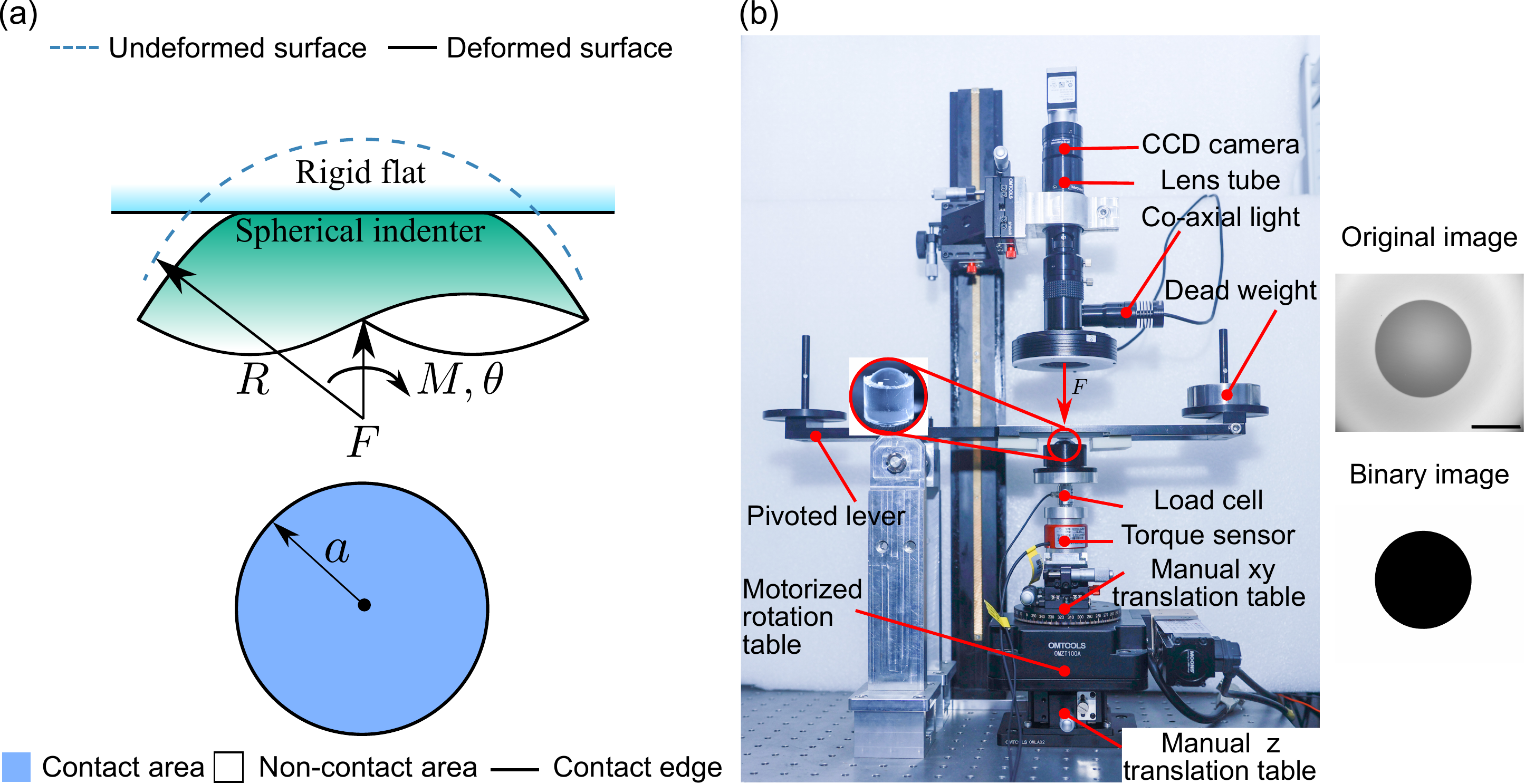}
  \caption{(a) Sketch of the torsional contact problem; (b) Left: the opto-mechanical rig used in the experiments. Right: typical contact image (scale bar: $3$ mm) and corresponding segmented image, enabling measurement of the contact area. Reproduced with permission from \cite{zhang2024non}, Copyright 2024 Springer.}\label{fig:Fig_1}
\end{figure}

Fig. \ref{fig:Fig_1}(a) illustrates the torsional contact pair consisting of a PDMS spherical surface with a radius of curvature $R = 13.11$ mm and a flat glass plate. The glass plate is first pressed against the PDMS surface under a constant normal force $F \in [2.57~\text{N}, 9.71~\text{N}]$ and held for a dwell period until the contact area is stabilized. The contact pair is then subjected to an increasing twist angle $\theta \in [0^{\circ}, 100^{\circ}]$ driven by a motorized rotation table at a constant angular velocity $\dot{\theta} \in [0.1^{\circ}/\text{s}, 12^{\circ}/\text{s}]$, such that the torsional loading remains quasi-static. The contact area formed on the interface is a circle of the radius $a$. 

In the present study, the torsional contact test is performed using a customized opto-mechanical device as illustrated in Fig. \ref{fig:Fig_1}(b). The normal force is the mean measured value by the load cell throughout the entire torsional test, which is achieved by applying a dead weight through a lever. The twist angle is measured by a rotary encoder, while the angular velocity is controlled by a motorized rotation table. The torque $M$ is measured by a torque sensor, while the contact area $A$ is quantified optically. A monochrome CCD camera (acA2440-20gm, Basler) with a $2448 \times 2048$ pixel sensor and 10-bit depth was positioned above the transparent glass counter-surface to record the contact region at 5 frames per second. The spatial resolution of the imaging system is approximately $4.7~\mu\text{m}$ per pixel. The spatial resolution is calibrated using a negative 1951 USAF resolution test target (RTS3AB-N, LBTEK, China). No image filtering was applied to the raw images before the contact area was thresholded using Otsu's method. Due to the sharp optical contrast between the contacting and non-contacting regions, the edge-detection ambiguity is $1$--$2$ pixels (approximately $8$--$16~\mu\text{m}$). In each frame, the outer boundary of the contact patch was identified automatically via intensity thresholding, taking advantage of the sharp optical contrast between the contacting and non-contacting regions. Pixels falling within the detected boundary were assigned to the contact zone, while those outside were treated as background, yielding a binary image from which the contact area was computed. More details of the rig and the PDMS fabrication procedure can be found in Zhang et al. \cite{zhang2024non}. Without further specification, all curves shown in Sections \ref{sec:Oscillating_contact} and \ref{sec:Large_deformation} are the mean over five independent samples, and error bands represent $\pm$ one standard deviation. In figures where multiple curves exhibit error bands of comparable magnitude, only one representative error band is displayed for clarity.

\begin{figure}[!ht]
  \centering
  \includegraphics[width=16cm]{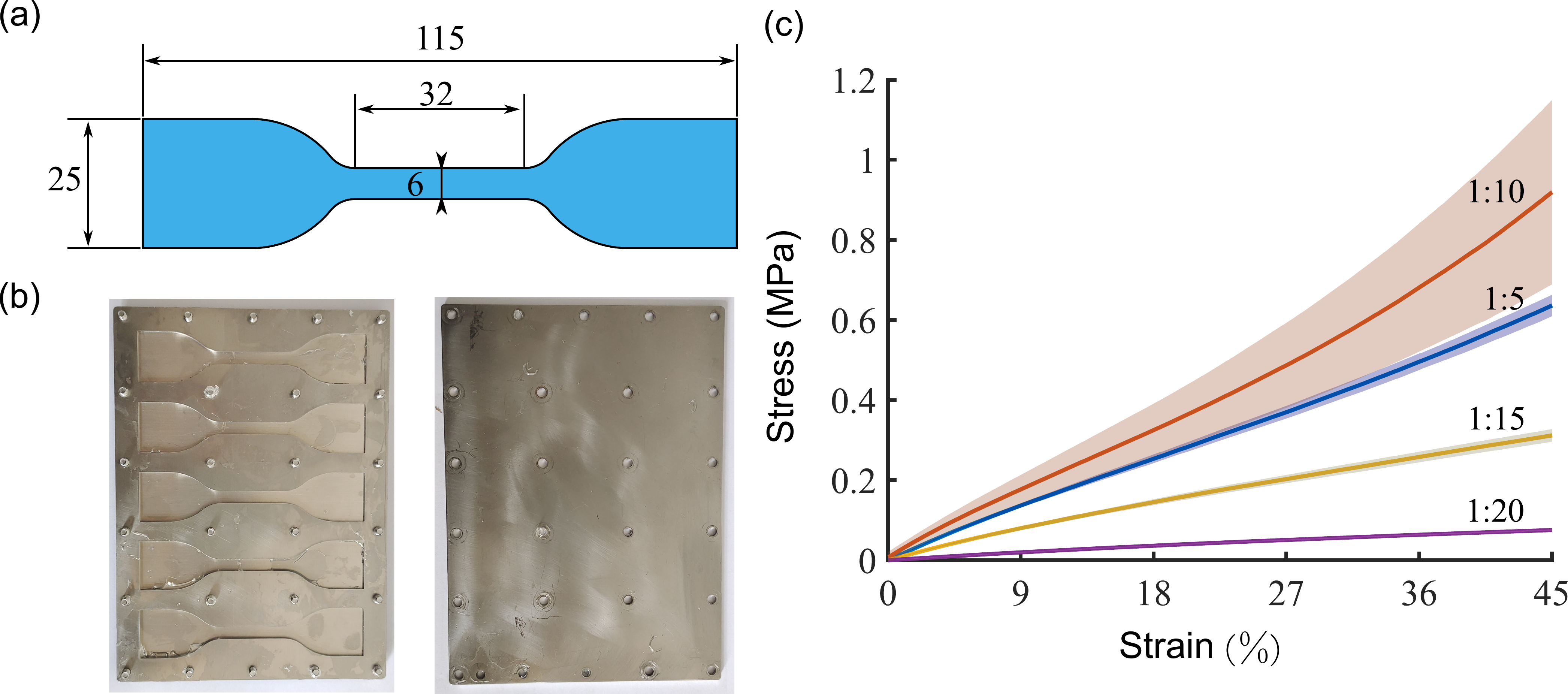}
  \caption{(a) Dimensions of specimen for uniaxial tensile test (unit: mm); (b) Bottom (left) and top (right) steel mold for specimen fabrication; (c) Engineering stress-strain curves of PDMS material with four different mixing ratios. Colored regions in (c) are error bands ($\pm$ standard deviation) over the six PDMS samples.}\label{fig:Fig_6}
\end{figure}

Four groups of PDMS spherical samples were fabricated with four curing-agent-to-base ratios (mixing ratios), namely $1:5$, $1:10$, $1:15$, and $1:20$, with five spherical samples in each group. Mixing ratios lower than $1:20$ were not considered because the resulting PDMS surfaces were too soft and adhesive to be used reliably in the torsional contact test, where severe adhesive wear could occur \cite{venkatadri2023torsion}. The mechanical response of PDMS with different mixing ratios was characterized by quasi-static uniaxial tensile tests using a universal material testing machine (TSKL, Tinius Kuli, China) at a crosshead speed of $5$ mm/min. The specimen dimensions are shown in Fig. \ref{fig:Fig_6}(a). The tensile specimens were fabricated using the same PDMS preparation procedure as that used for the spherical samples \cite{zhang2024non}, except that a different mold was employed, as shown in Fig. \ref{fig:Fig_6}(b). 

The Poisson's ratio of the PDMS sample is assumed to be 0.5. The mean engineering stress-strain curves obtained from six replicas are presented in Fig. \ref{fig:Fig_6}(c). The dispersion among the six replicas is generally small, except for the $1:10$ samples. In the tensile tests, the maximum strain was limited to below $45 \%$ in order to avoid edge-initiated fracture or slippage between the specimen and the grips. The non-monotonic variation of Young's modulus with the mixing ratio has been observed repeatedly in the previous studies \cite{Johnston2019PDMSModulus, Cho2021PdmsYoungModulus}. The Young's modulus is strongly correlated with the crosslink density, which is maximized when the base and curing agent are at stoichiometric balance. Either an excess of base or curing agent from the balance point will lead to a decrease in crosslink density \cite{Lee2016EffectOfCuringAgentConcentrationPDMS}.

\section{Oscillating torsional contact}\label{sec:Oscillating_contact}
The reversibility of torsional contact is examined through one complete oscillating cycle consisting of loading (forward twist) followed by unloading (backward twist). Before torsional loading, the contact pair is subjected to pure normal loading for $2$ min. The loading stage is then initiated by increasing the twist angle from $0^{\circ}$ to $100^{\circ}$ at a constant angular velocity. After the contact pair is held stationary at $100^{\circ}$ for another $2$ min, the unloading stage is initiated at the same angular velocity until the twist angle returns to zero. The normal force is maintained by deadweight throughout the entire loading-unloading cycle. The mixing ratio of the PDMS surface is fixed at $1:10$ in this section. 

\begin{figure}[!ht]
  \centering
  \includegraphics[width=14cm]{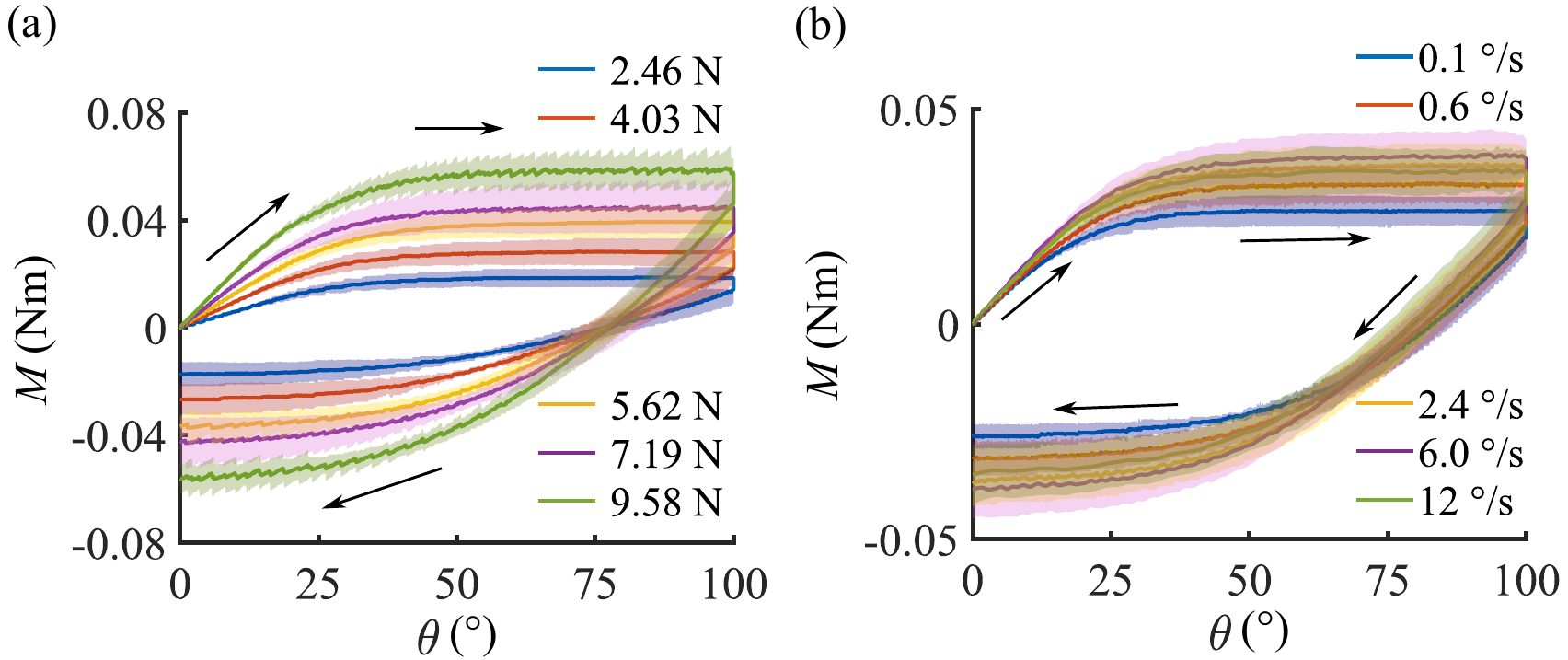}
  \caption{Variation of torque with respect to twist angle in a complete loading-unloading torsional cycle under (a) various normal forces and (b) various values of angular velocity. Black arrows indicate the loading/unloading direction. The angular velocity in (a) and normal force in (b) are 0.6$^{\circ}$/s and 5.62 N, respectively.}\label{fig:Fig_2}
\end{figure}

Fig. \ref{fig:Fig_2} shows that the variation of $M$ with $\theta$ during a complete loading-unloading cycle forms a partial hysteresis loop. During the loading stage, where the twist angle increases monotonically, $M$ initially increases linearly with $\theta$ during partial slip and then gradually approaches a constant value once the sliding stage is reached. The twist angle is subsequently held at $\theta = 100^{\circ}$ for $2$ min, during which the torque decreases because of creep deformation. During the unloading stage, where the twist angle decreases monotonically, $M$ decreases linearly with decreasing $\theta$ and then gradually approaches a constant value with the opposite sign. Notably, the unloading curves of $M(\theta)$ intersect near a common point with $M_{\text{i}} \approx 0$ Nm and $\theta_{\text{i}} \approx 75^{\circ}$ for different normal forces (Fig. \ref{fig:Fig_2}(a)) and $\theta_{\text{i}} \approx 70^{\circ}$ for different angular velocities (Fig. \ref{fig:Fig_2}(b)). Neglecting the dwell time portion, these hysteresis loops are in qualitative agreement with Deresiewicz's torsional contact model \cite{deresiewicz1954contact, segalman2005new}. The irreversibility arises because, at the beginning of the unloading stage, the circumferential shear stress in the outer slip region reverses sign. As the normal force and angular velocity increase, the torque in the sliding stage increases, resulting in an expansion of the hysteresis loop.

\begin{figure}[!ht]
  \centering
  \includegraphics[width=14cm]{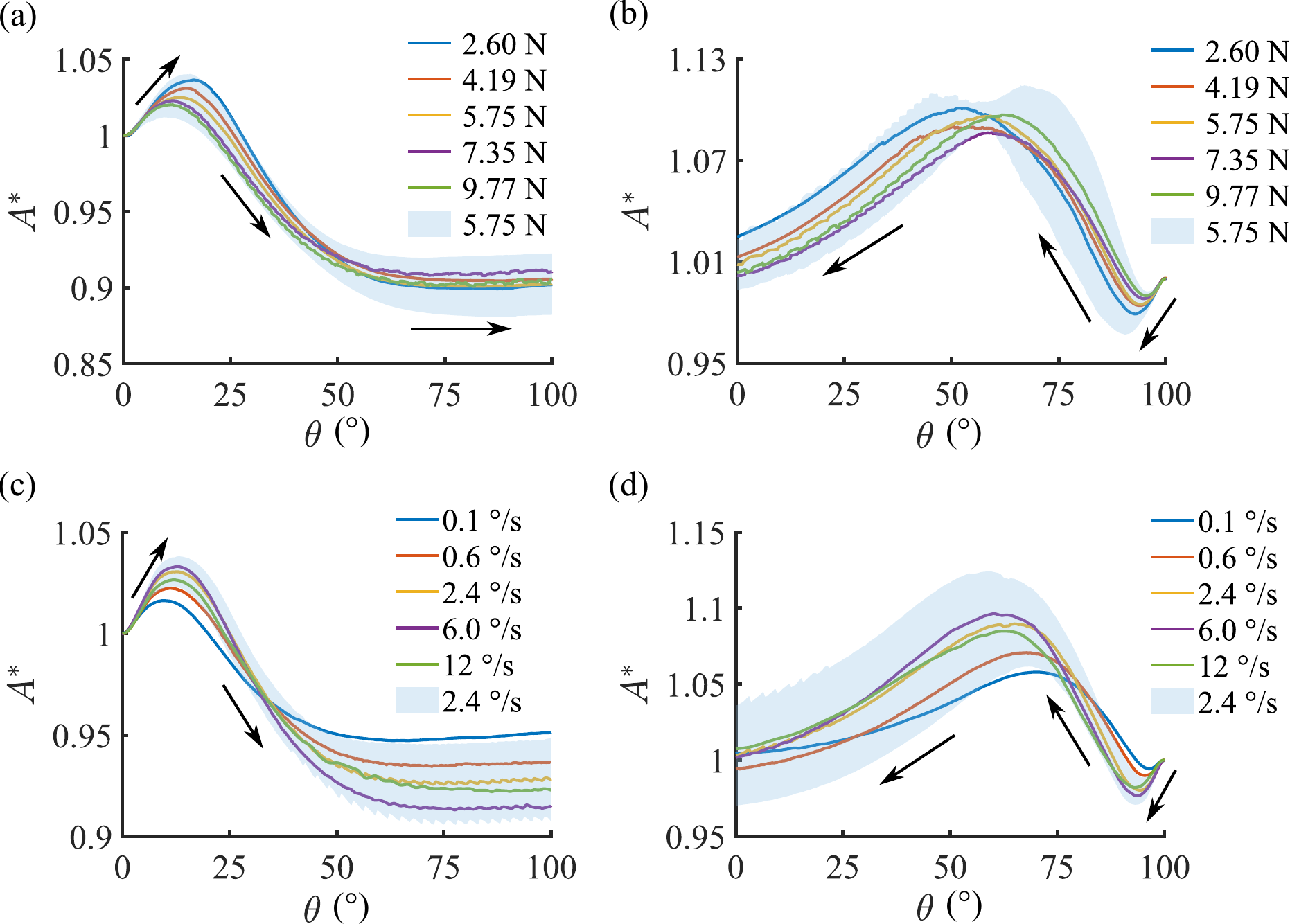}
  \caption{Variation of normalized contact area ($A^*$) with respect to twist angle: (a, c) Contact pair is subjected to various normal forces and angular velocities in the loading stage where $A^* = A(\theta)/A(\theta = 0^{\circ})$; (b, d) Contact pair is subjected to various normal forces and angular velocities in the unloading stage where $A^* = A(\theta)/A(\theta = 100^{\circ})$ and $A(\theta = 100^{\circ})$ is the initial contact area in the unloading stage. The angular velocity in (a, b) and normal force in (c, d) are 0.6$^{\circ}$/s and 5.75 N, respectively.}\label{fig:Fig_3}
\end{figure}

During the loading stage, the normalized contact area $A^*$ varies non-monotonically, first increasing and then decreasing smoothly until a nearly constant value is reached once the torsional interface enters full sliding (Fig. \ref{fig:Fig_3}(a, c)). This is exactly the same non-monotonic trend as reported in Zhang et al. \cite{zhang2024non}. For the initial contact area increase observed at the onset of torsional loading, the peak relative area change ($\Delta A^*_{\text{peak}} \approx 2\%$-$5\%$) is small compared with the subsequent area reduction. Since the light intensity contrast is sharp at the contact edge, the edge-detection ambiguity is $1$-$2$ pixels, corresponding to a relative area uncertainty significantly lower than $1\%$. Therefore, the initial area increase is a robust feature of the contact evolution, despite the small magnitude.

At the beginning of the unloading stage, the variation of $A^*$ with $\theta$ becomes even more complex: $A^*$ first decreases slightly, then increases to a maximum in the middle of the unloading stage, and finally decreases again toward the end of unloading. Within each result group in Fig. \ref{fig:Fig_3}, the error bands of the $A^*(\theta)$ curves are approximately the same, so only one representative error band is shown in each panel. For a fixed twist angle, the dispersion of $A^*(\theta)$ is negligible during the initial increase in the loading stage (Fig. \ref{fig:Fig_3}(a, c)) and during the initial decrease in the unloading stage (Fig. \ref{fig:Fig_3}(b, d)). At later stages, however, the dispersion increases abruptly by more than one order of magnitude, which is attributed to the stick-slip-like oscillation of the contact area \cite{zhang2024non}. The changes in $A^*(\theta)$ induced by different normal forces (Fig. \ref{fig:Fig_3}(b)) and angular velocities (Fig. \ref{fig:Fig_3}(d)) are comparable to, or even larger than, the sample-to-sample dispersion. Therefore, $A^*(\theta)$ during the loading-unloading cycle depends only weakly on the normal force and angular velocity. 

\begin{figure}[!ht]
  \centering
  \includegraphics[width=12cm]{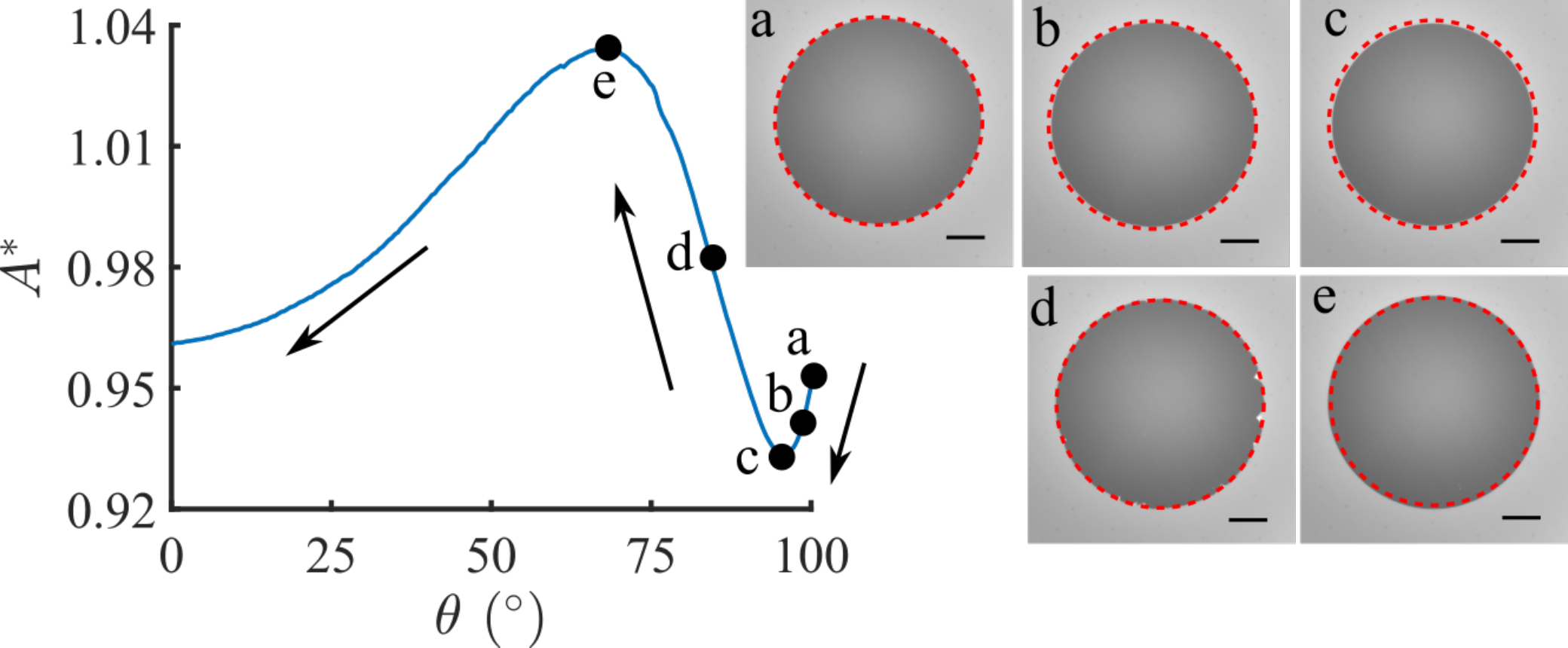}
  \caption{A typical variation of normalized contact area with respect to twist angle during unloading stage. Five contact area images are captured at five representative instants: (a) $100^{\circ}$, (b) $97^{\circ}$, (c) $95^{\circ}$, (d) $82^{\circ}$, (e) $70^{\circ}$. The red dashed line represents the boundary of the contact area in image a. Scale bar: 1 mm. Normal force and angular velocity are $5.62$ N and $0.6^{\circ}$/s.}\label{fig:Fig_4}
\end{figure}

\begin{figure}[!ht]
  \centering
  \includegraphics[width=11cm]{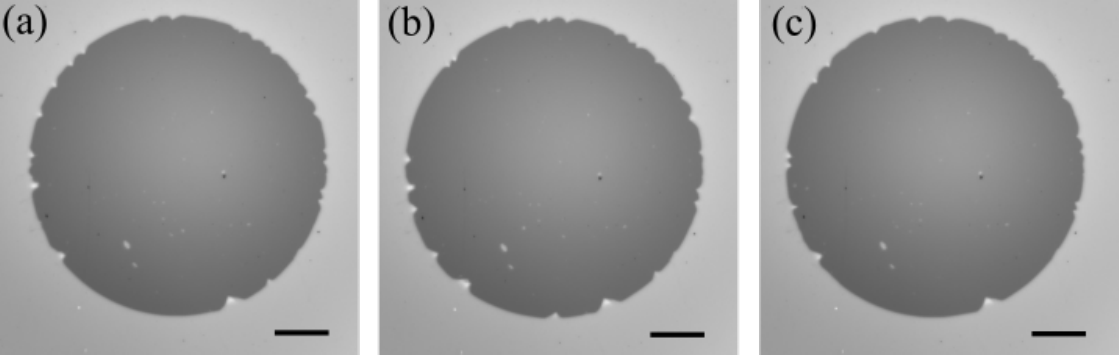}
  \caption{Three images of the contact area of three different PDMS samples captured at the same twist angle ($\theta = 78^{\circ}$) during the unloading stage. Scale bar: 1 mm. Normal force and angular velocity are $5.62$ N and $0.6^{\circ}$/s.}\label{fig:Fig_5}
\end{figure}

To further examine the variation of contact area during the unloading stage, contact area images are shown at five representative moments. At the first three moments, all within the initial decrease stage, the contact area remains nearly circular (see insets a, b, and c in Fig. \ref{fig:Fig_4}). Petal-like edges, induced by the surface wrinkling, begin to appear at approximately $\theta = 87^{\circ}$ (see inset d in Fig. \ref{fig:Fig_4} as an example) only when the angular velocity is sufficiently high ($\dot{\theta} > 0.6^{\circ}$/s). Interestingly, the petal-like morphology is highly repeatable among different samples at the same twist angle (Fig. \ref{fig:Fig_5}). This observation indicates that the twist-induced wrinkling is not caused by the random roughness, but is instead an intrinsic surface instability of the soft elastomer. After the contact area decreases from the maximum value at moment e, its morphology returns to a circle with a smooth circular edge. Given that the same binarization threshold is applied consistently across all frames and samples, the edge-detection ambiguity is $1$-$2$ pixels (approximately $8$-$16~\mu\text{m}$), which does not qualitatively affect the majority of the observed trends of contact area evolution whose petal-like feature size is larger than the size uncertainty (see the petal-like feature in Figs. \ref{fig:Fig_4}(d) and \ref{fig:Fig_5}).

\section{Large deformation}\label{sec:Large_deformation}
Motivated by previous finite element results \cite{lengiewicz2020finite,ceglie2025contact,mergel2021contact} suggesting that large deformation may drive shear-induced contact area reduction, we experimentally test this hypothesis by systematically varying the softness of the PDMS surface through the mixing ratio. In parallel, whether the large deformation also drives the initial contact area increase is also investigated. As shown in Fig. \ref{fig:Fig_6}, the PDMS surfaces with mixing ratios of $1:15$ and $1:20$ are softer than those with $1:5$ and $1:10$, and thus exhibit relatively larger deformation under the same torsional loading. 

\begin{figure}[!ht]
  \centering
  \includegraphics[width=13cm]{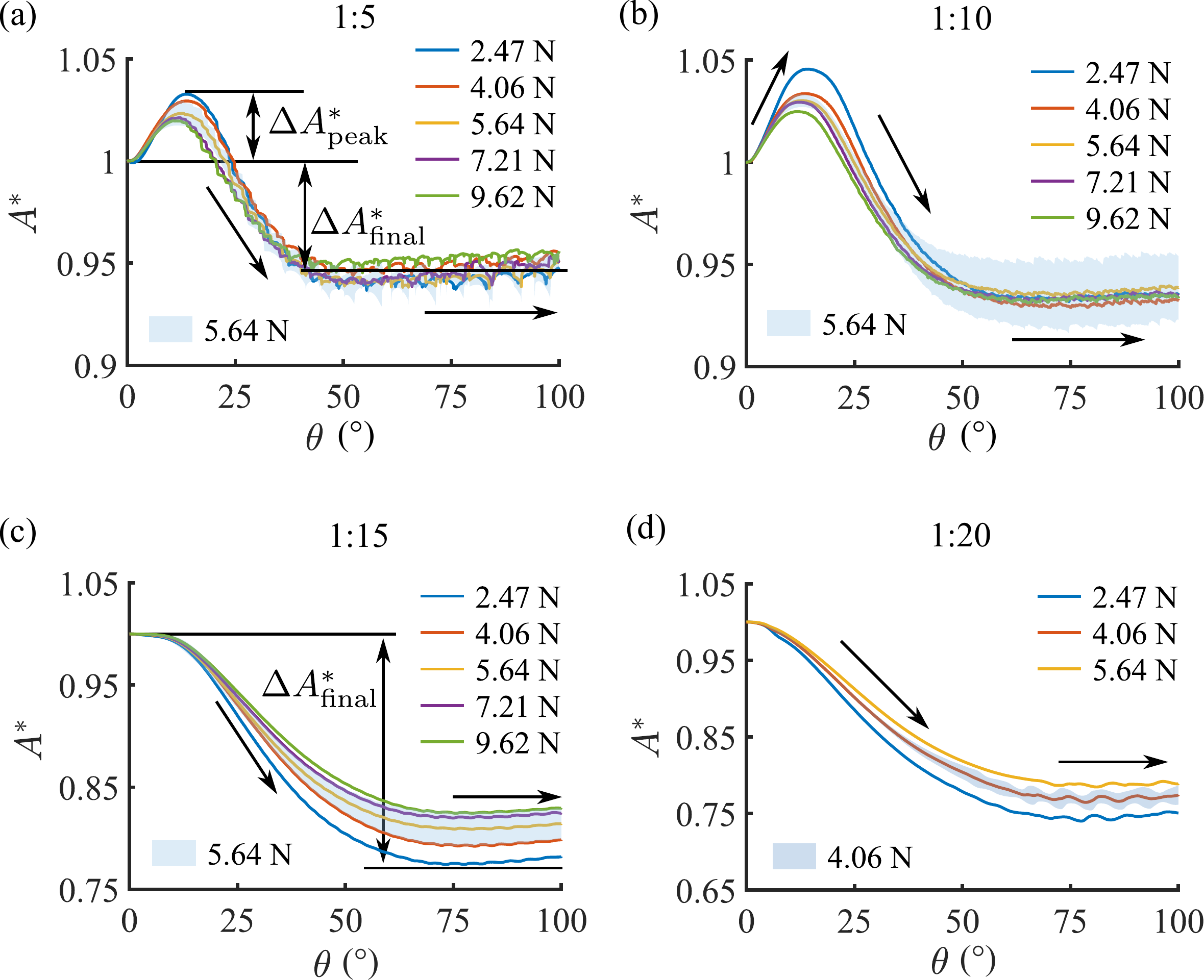}
  \caption{Variations of normalized contact area with twist angle at the loading stage under various normal forces. Mixing ratios are (a) $1:5$, (b) $1:10$, (c) $1:15$, (d) $1:20$. $A^* = A(\theta)/A(\theta = 0^{\circ})$. Black arrows indicate the loading direction. Only three normal forces are used with $1:20$ since the contact area expands beyond the field of view of the camera when $F > 5.64$ N. Angular velocity is $0.6^{\circ}$/s.}\label{fig:Fig_7}
\end{figure}

\begin{figure}[!ht]
  \centering
  \includegraphics[width=13cm]{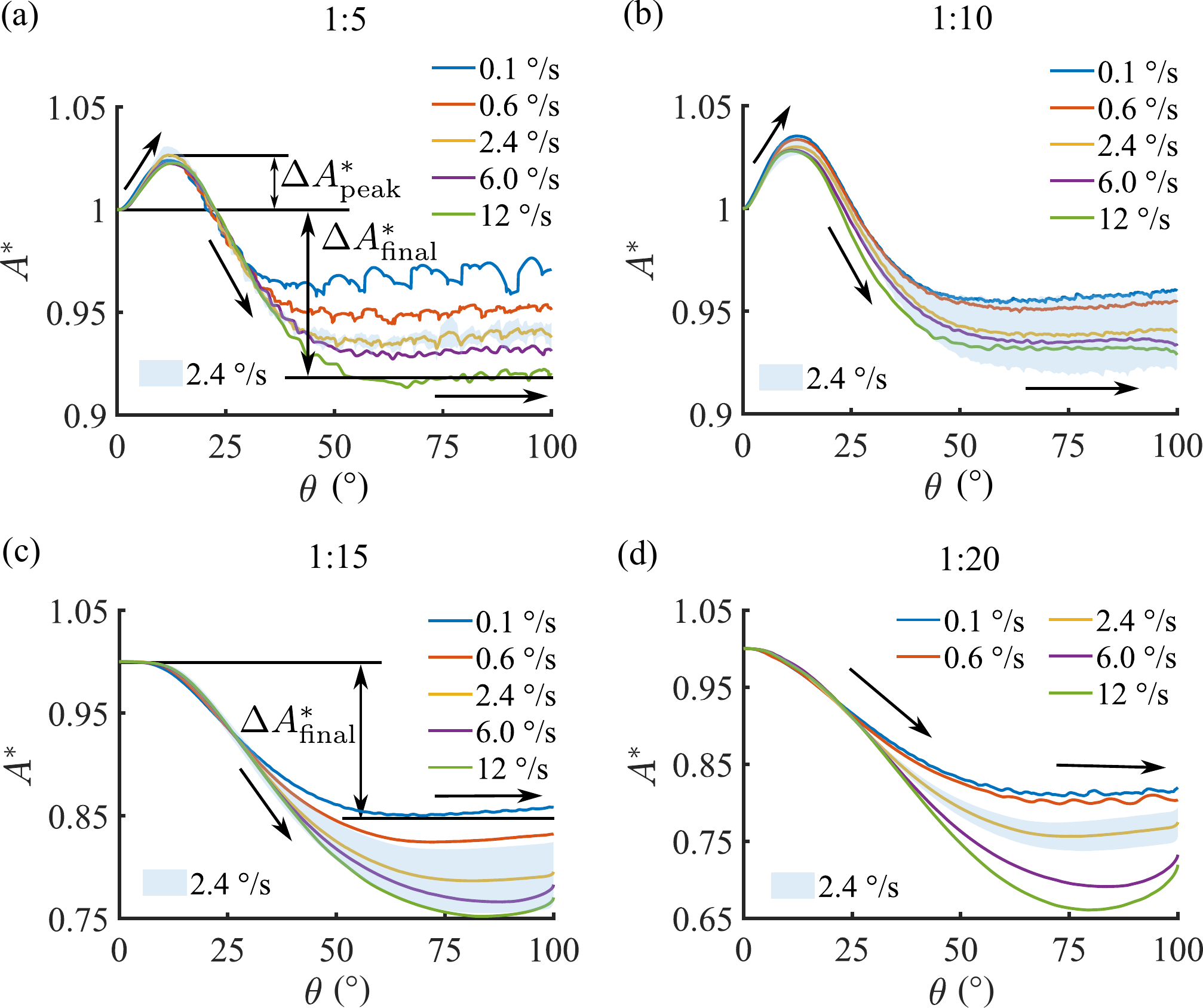}
  \caption{Variations of normalized contact area $A^* = A(\theta)/A(\theta = 0^{\circ})$ with $\theta$ at the loading stage with various values of angular velocity. Mixing ratios are (a) $1:5$, (b) $1:10$, (c) $1:15$, (d) $1:20$. Black arrows indicate the loading direction. Normal force is $5.64$ N.}\label{fig:Fig_8}
\end{figure}

The effect of the mixing ratio on $A(\theta)$ at the loading stage under various normal forces is explored in Fig. \ref{fig:Fig_7}. When the mixing ratio is $1:5$ or $1:10$, the variation of the relative contact area $A^* = A(\theta)/A(\theta = 0^{\circ})$ with $\theta$ exhibits a non-monotonic trend, first increasing by $\Delta A_{\text{peak}}^*$ (see Fig. \ref{fig:Fig_7}(a)) and then followed by a decreasing trend until a final drop of $\Delta A_{\text{final}}^*$ is reached. According to Fig. \ref{fig:Fig_6}(c), PDMS ($1:5$) is softer than PDMS ($1:10$). The value of $\Delta A_{\text{peak}}^*$ of softer PDMS ($1:5$) is slightly lower than that of $1:10$. Notably, the initial increasing trend of $A(\theta)$ disappears ($\Delta A_{\text{peak}}^* = 0$) once the mixing ratio of the PDMS surface drops to $1:15$ and $1:20$. Moreover, the contact area drops more rapidly with $\theta$ at lower mixing ratios ($\Delta A_{\text{final}}^* \approx 5\%$ at $1:5$ and $\Delta A_{\text{final}}^* \approx 25\%$ at $1:20$). The variation of $A^*(\theta)$ shows only minor dependence on the normal force when the mixing ratio is relatively high ($1:5$ and $1:10$). As the mixing ratio further drops to $1:15$ and $1:20$, the reduction of $A^*$ with $\theta$ becomes more pronounced under a smaller normal force. Similar tests with various angular velocities during the loading stage were conducted, with the results shown in Fig. \ref{fig:Fig_8}, where the normal force is kept constant. The initial area increasing trend remains absent regardless of the selected angular velocities. The area reduction is also enhanced with increasing angular velocity. 

\begin{figure}[!ht]
  \centering
  \includegraphics[width=13cm]{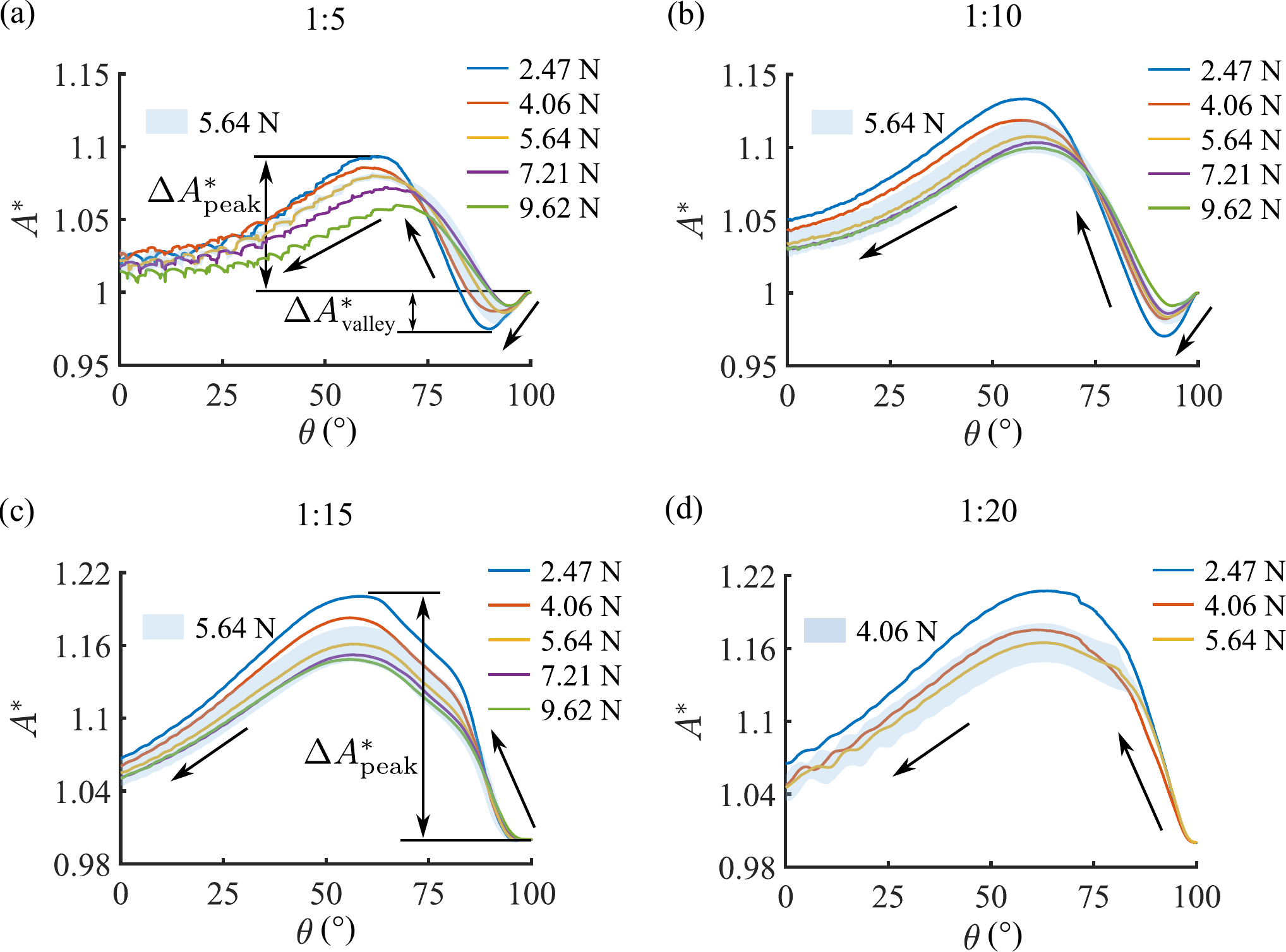}
  \caption{Variations of normalized contact area $A^* = A(\theta)/A(\theta = 100^{\circ})$ with $\theta$ at the unloading stage under various normal forces, where $A(\theta = 100^{\circ})$ is the initial contact area at the beginning of the unloading stage after 2-min dwell time. Mixing ratios are (a) $1:5$, (b) $1:10$, (c) $1:15$, (d) $1:20$. Black arrows indicate the unloading direction. Angular velocity is $0.6^{\circ}$/s.}\label{fig:Fig_9}
\end{figure}

\begin{figure}[!ht]
  \centering
  \includegraphics[width=13cm]{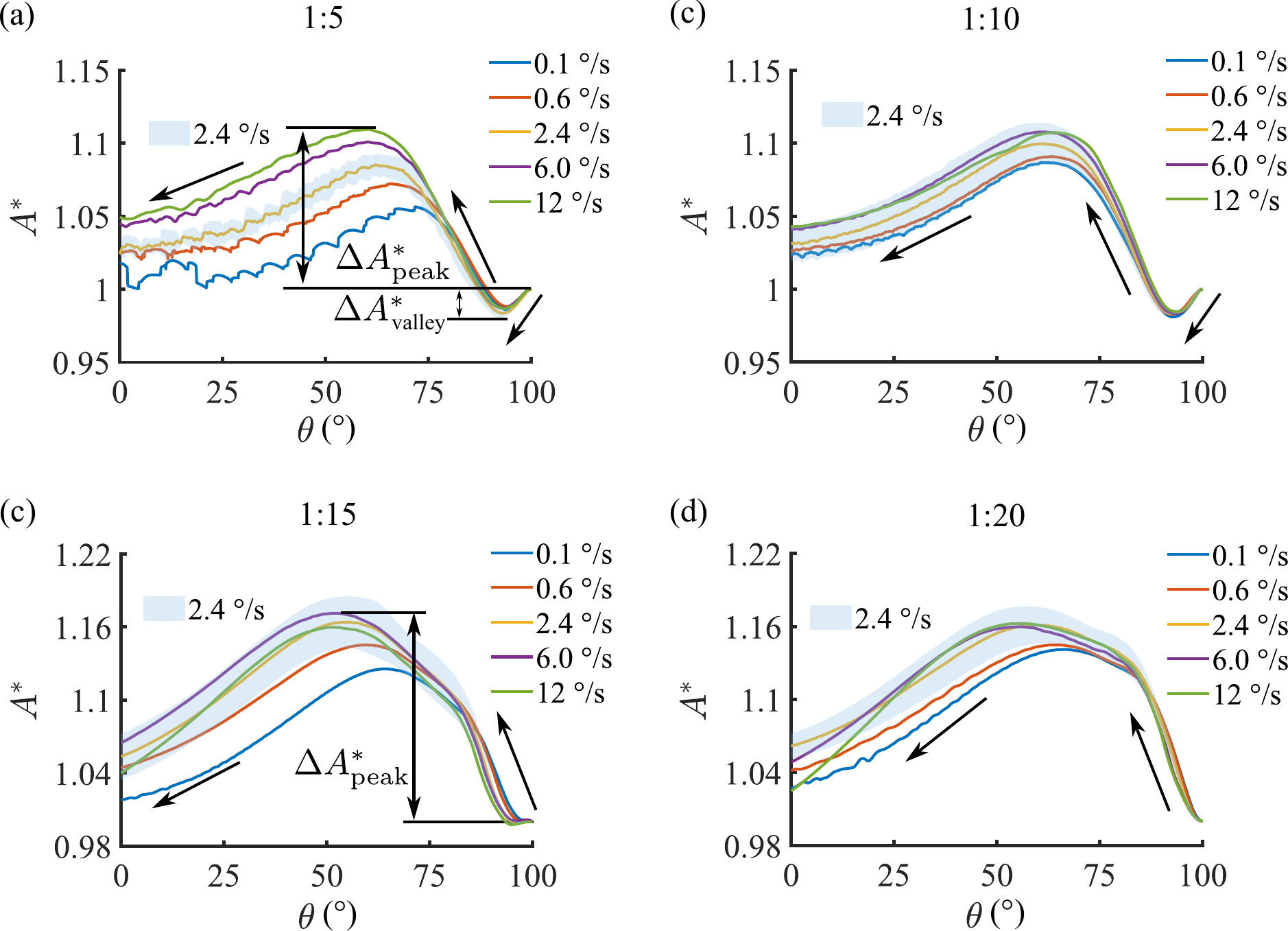}
  \caption{Variations of normalized contact area $A^* = A(\theta)/A(\theta = 100^{\circ})$ with $\theta$ during unloading with various values of angular velocity, where $A(\theta = 100^{\circ})$ is the initial contact area at the beginning of the unloading stage after 2-min dwell time. Mixing ratios are (a) $1:5$, (b) $1:10$, (c) $1:15$, (d) $1:20$. Black arrows indicate the unloading direction. Normal force is $5.64$ N.}\label{fig:Fig_10}
\end{figure}

\begin{figure}[!ht]
  \centering
  \includegraphics[width=13cm]{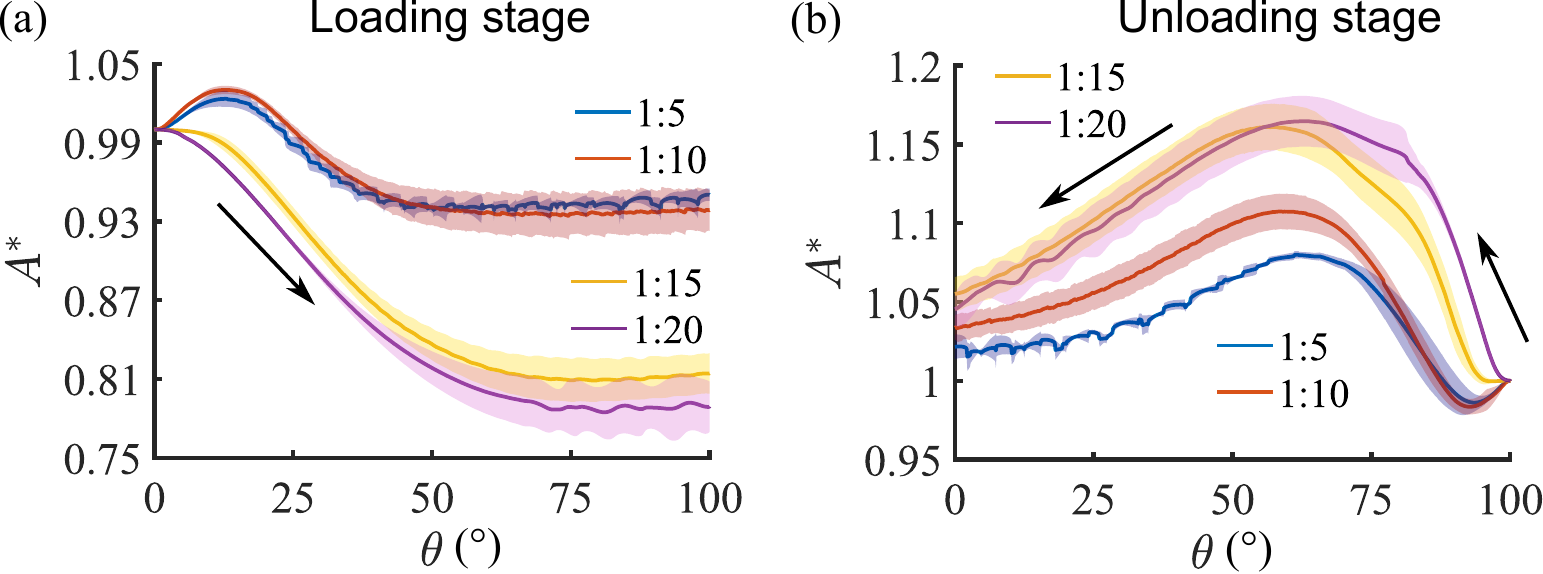}
  \caption{Variation of normalized contact area with $\theta$ during the (a) loading and (b) unloading stages, where $A^* = A(\theta)/A(\theta = 0^{\circ})$ and $A^* = A(\theta)/A(\theta = 100^{\circ})$, respectively. The normal force is $5.64$ N and angular velocity is $0.6^{\circ}$/s. Black arrows in (a) and (b) indicate the loading and unloading directions, respectively.}\label{fig:Fig_13}
\end{figure}

Figs. \ref{fig:Fig_9} and \ref{fig:Fig_10} show a more complex variation of the contact area during the unloading stage. The twist angle is kept at $100^{\circ}$ for $2$ min before unloading begins. The relative contact area at the unloading stage is defined as $A^*(\theta) = A(\theta)/A(\theta = 100^{\circ})$ for $\theta < 100^{\circ}$, where $A(\theta = 100^{\circ})$ is the contact area at the start of unloading. When the mixing ratio is $1:5$ and $1:10$, the $A^*(\theta)$ relation shows a decrease-increase-decrease pattern similar to that shown in Section 3. The maximum drop in the first decreasing trend ($\Delta A^*_{\text{valley}}$ as marked in Figs. \ref{fig:Fig_9} and \ref{fig:Fig_10}) is relatively insensitive to the normal force and angular velocity according to error bands. When the mixing ratio decreases to $1:15$ and $1:20$, the initial decreasing trends disappear. The maximum $A^*$ ($\Delta A^*_{\text{peak}}$ as marked in Figs. \ref{fig:Fig_9} and \ref{fig:Fig_10}) stays almost the same as the mixing ratio decreases. The maximum increase in contact area during unloading (i.e., $\Delta A^*_{\text{valley}} + \Delta A^*_{\text{peak}}$) significantly exceeds that during loading, a disparity that is further amplified by the high compliance of the PDMS material. To better illustrate the influence of the mixing ratio on the variation of the contact area, we compare $A(\theta)$ curves associated with different mixing ratios and a typical combination of normal force and angular velocity in Fig. \ref{fig:Fig_13}. 

\begin{figure}[!ht]
  \centering
  \includegraphics[width=13cm]{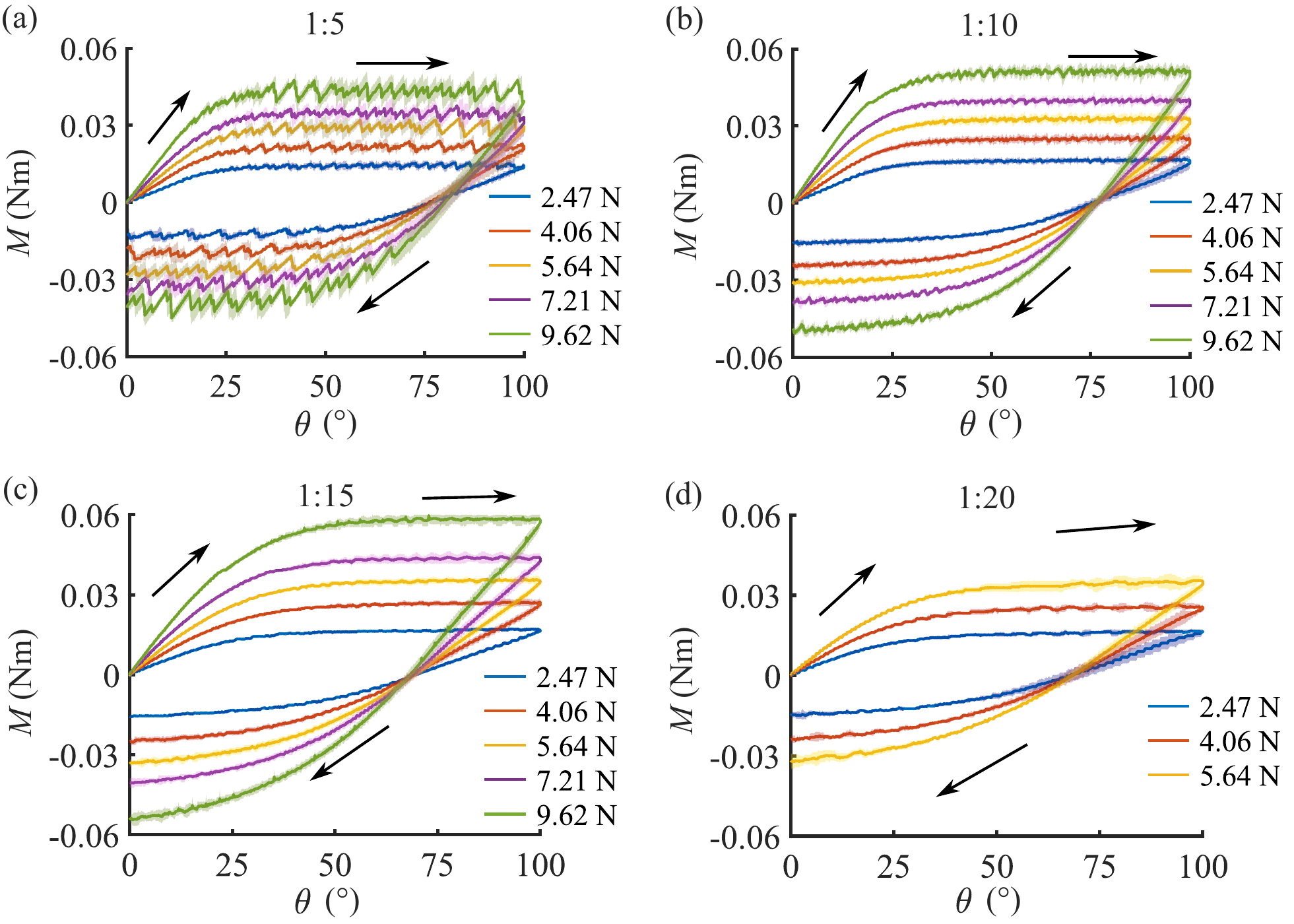}
  \caption{Variations of torque with twist angle during a complete loading-unloading cycle under various normal forces. Mixing ratios are (a) $1:5$, (b) $1:10$, (c) $1:15$, (d) $1:20$. Black arrows indicate the loading/unloading direction. Angular velocity is $0.6^{\circ}$/s.}\label{fig:Fig_11}
\end{figure}

\begin{figure}[!ht]
  \centering
  \includegraphics[width=13cm]{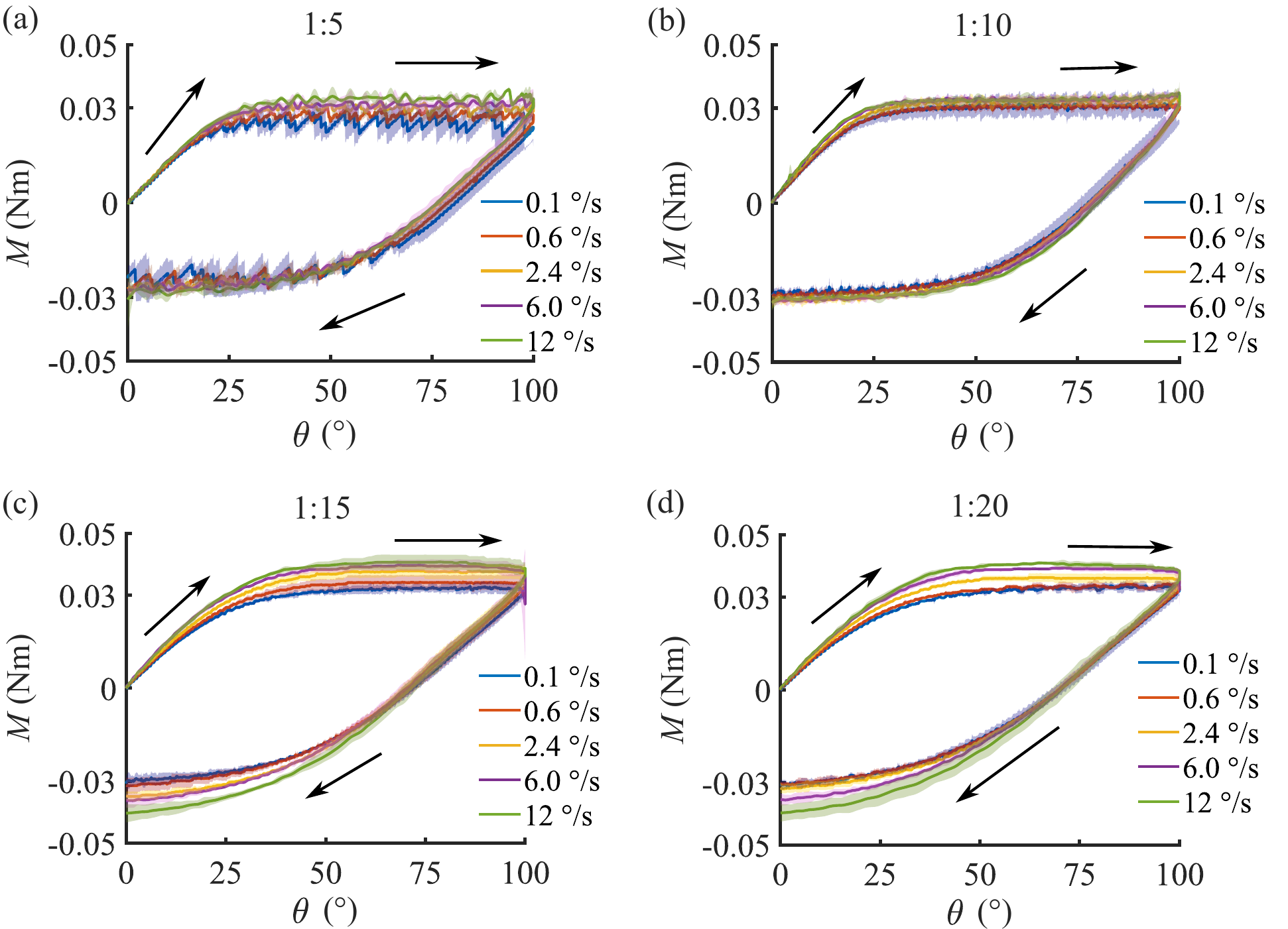}
  \caption{Variations of torque with twist angle during a complete loading-unloading cycle with various values of angular velocity. Mixing ratios are (a) $1:5$, (b) $1:10$, (c) $1:15$, (d) $1:20$. Black arrows indicate the loading/unloading direction. Normal force is $5.64$ N.}\label{fig:Fig_12}
\end{figure}

Figs. \ref{fig:Fig_11} and \ref{fig:Fig_12} show that as the mixing ratio decreases, the hysteresis loop formed by $M(\theta)$ slightly expands vertically, with a higher maximum $M$. The oscillated torque at the plateau portion of $A(\theta)$ during both loading and unloading stages, with a relatively high mixing ratio ($1:5$ and $1:10$), is greatly relieved with a low mixing ratio ($1:15$ and $1:20$), regardless of the normal force and angular velocity. As a matter of fact, the repeatability of $M(\theta)$ curves is significantly improved with lower mixing ratios. The intersection points of unloading curves in Fig. \ref{fig:Fig_11} remain in the vicinity of ($M_{\text{i}} = 0$ Nm, $\theta_{\text{i}} = 75^{\circ}$), while $M_{\text{i}}$ increases towards the negative direction and $\theta_{\text{i}}$ decreases as the mixing ratio decreases. Unlike the strong influence of the normal force on the hysteresis loop of $M(\theta)$, angular velocity has a relatively weaker effect on $M(\theta)$. The influence becomes slightly more evident with the mixing ratio of $1:15$ and $1:20$ (Fig. \ref{fig:Fig_12}). 

\begin{figure}[H]
  \centering
  \includegraphics[width=13cm]{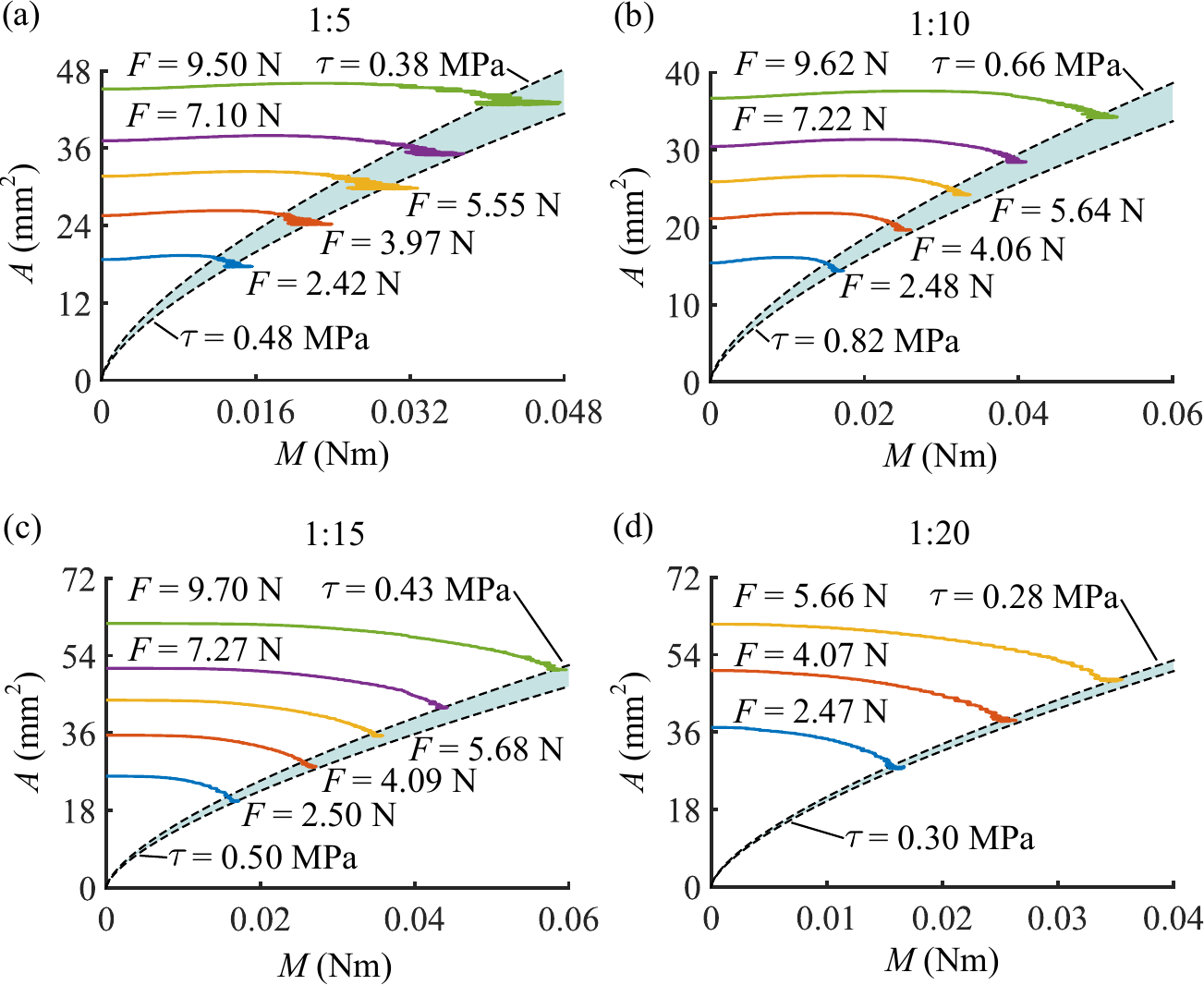}
  \caption{Variation of contact area with torque during loading stage under various normal forces; The lateral oscillations at the right side of all solid lines are due to oscillations of the torque during macroscopic sliding. Bottom/top dashed lines and blue shaded area correspond respectively to the plot of Eq. \eqref{eq:AM_relation} with shear strength at the lower limit, upper limit, and values in between. The angular velocity is $0.6^{\circ}$/s.}\label{fig:Fig_14}
\end{figure}

Figure \ref{fig:Fig_14} illustrates the evolution of $A$ with respect to $M$ under different normal forces. When the mixing ratio is $1:5$ and $1:10$, the contact area varies monotonically until it reaches the sliding stage, where $M$ oscillates about its mean level. Assuming that the shear stress equals the shear strength ($\tau$) within the sliding zone, $A(M)$ theoretically terminates on the curve (denoted by dashed lines in Fig. \ref{fig:Fig_14}):
\begin{equation}\label{eq:AM_relation}
A = \pi^{1/3} \left( \frac{3}{2 \tau} \right)^{2/3} M^{2/3}. 
\end{equation}
Equation (1) assumes a uniform distribution of the shear stress, which is a reasonable approximation to the torque-area relation under fully sliding conditions \cite{zhang2024non}. For the partial-slip or stick-dominated stage, the uniform shear stress distribution is oversimplified and the value of $\tau$ represents only the mean shear stress over the contact area. 

\begin{figure}[H]
  \centering
  \includegraphics[width=13cm]{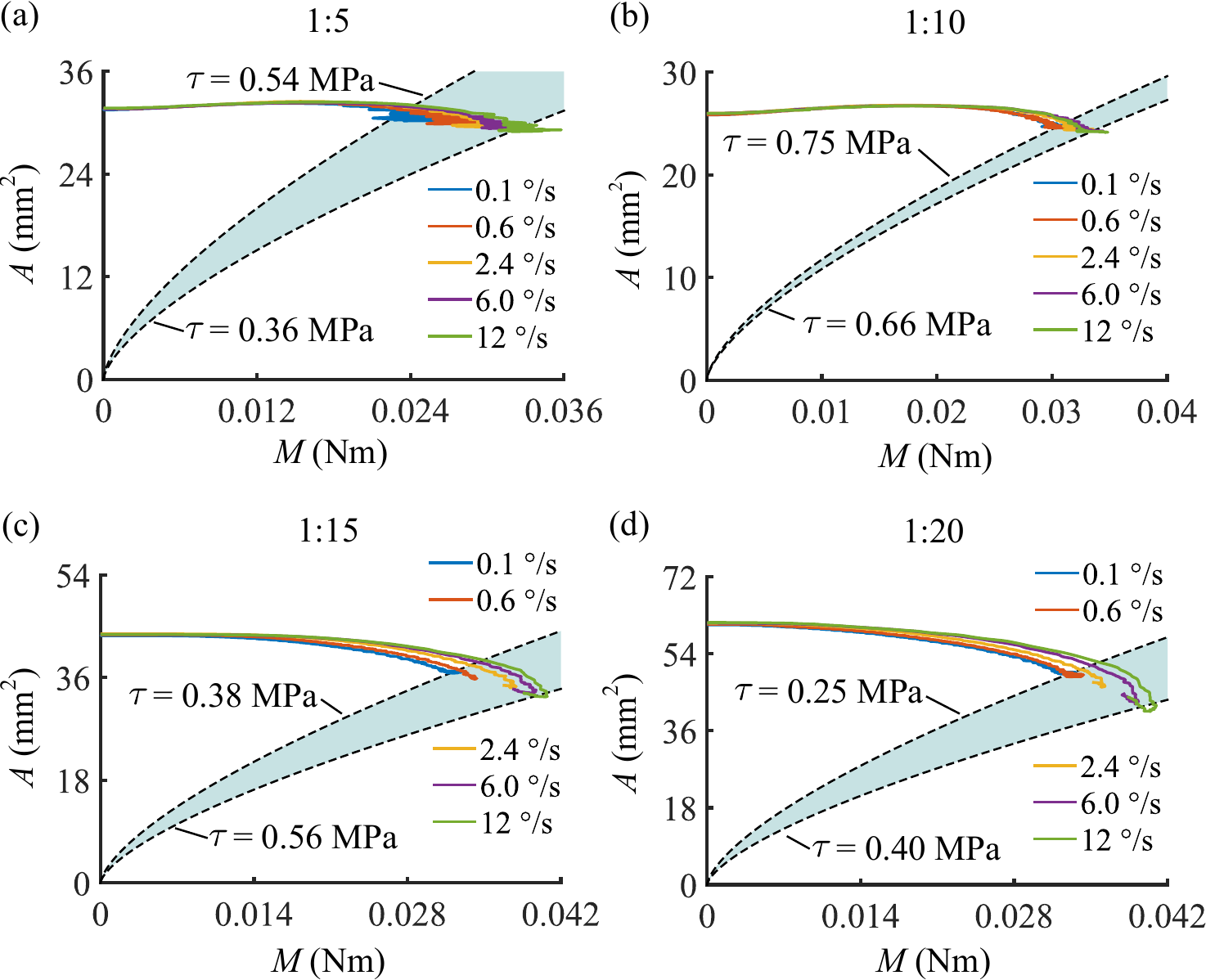}
  \caption{Variation of contact area with torque during loading stage with various values of angular velocity; The lateral oscillations at the right side of all solid lines are due to oscillations of the torque during macroscopic sliding. Bottom/top dashed lines and blue shaded area correspond respectively to the plot of Eq. \eqref{eq:AM_relation} with shear strength at the lower limit, upper limit, and values in between. The normal force is $5.64$ N.}\label{fig:Fig_15}
\end{figure}

\begin{figure}[H]
  \centering
  \includegraphics[width=13cm]{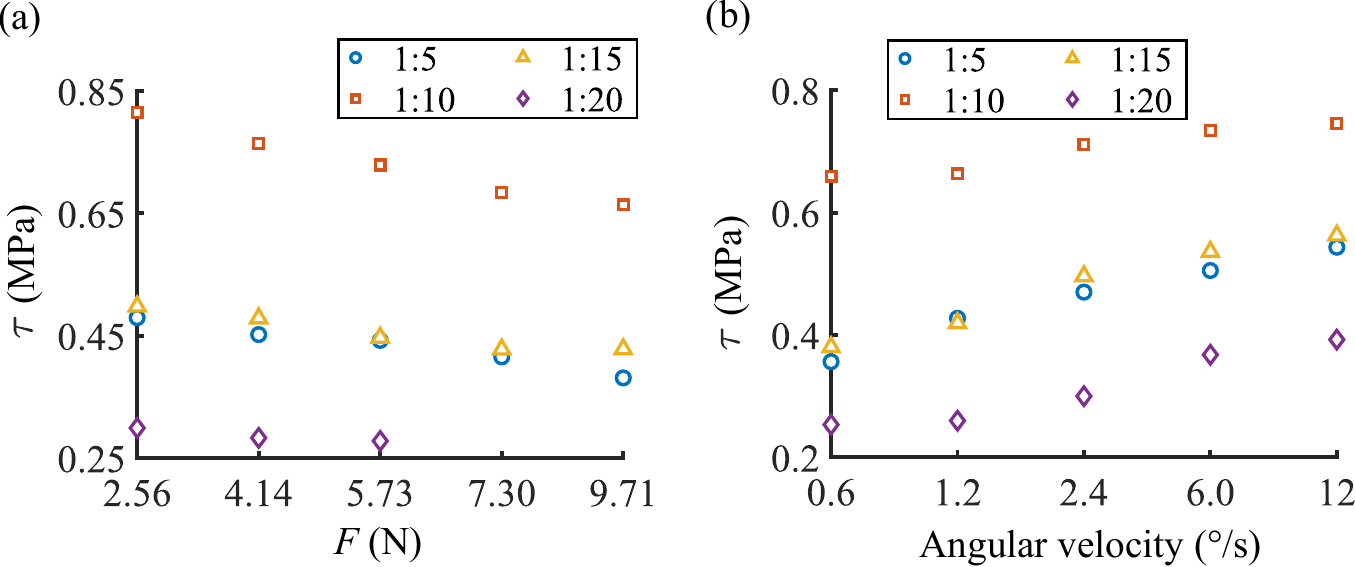}
  \caption{Variation of shear strength $\tau$ with (a) normal force and (b) angular velocity. The discrete data are obtained from Figs. \ref{fig:Fig_14} and \ref{fig:Fig_15}.}\label{fig:Fig_16}
\end{figure}

The shear strength is estimated from the test data by Eq. \eqref{eq:AM_relation}, where $M$ and $A$ are the mean torque and mean contact area in the slip state when $\theta > 40^{\circ}$. During each stick-slip-like cycle, the torque oscillates between a local maximum (static friction in stick state) and a local minimum (sliding friction in slip state) with an amplitude comparable with the mean torque. Therefore, the shear strength $\tau$ is sensitive to the selection of the sampling point within each cycle. Since Eq. \eqref{eq:AM_relation} assumes full sliding, the minimum-torque instant is used to estimate $\tau$ by back-calculating from the corresponding $M$ and $A$ values. Previous studies have shown that the shear strength increases with the normal force \cite{sahli2018evolution, zhang2024non}. This observation is again confirmed in Fig. \ref{fig:Fig_14}, with an additional finding that the shear strength is less sensitive to the normal force with lower mixing ratios (see the blue band enclosed by two dashed lines in Fig. \ref{fig:Fig_14}). The stiffer PDMS surfaces (with a mixing ratio of $1:5$ and $1:10$) have a relatively larger shear strength and variation than the softest PDMS (with a mixing ratio of $1:15$ and $1:20$). Figure \ref{fig:Fig_15} shows the evolution of $A(M)$ with angular velocity. As the mixing ratio decreases, the deviation of $A(M)$ curves with different angular velocities becomes more pronounced. A concise summary of how the shear strength varies with the normal force, angular velocity, and the mixing ratio is provided in Fig. \ref{fig:Fig_16}. 

\section{Discussions}\label{sec:Discussion}

The three novel aspects identified in the Introduction (loading-unloading torsional cycles, unloading-path irreversibility, and systematic variation of PDMS softness) collectively advance the understanding of soft torsional contacts beyond the prior work of Zhang et al. \cite{zhang2024non}. The loading-unloading torsional tests demonstrate a clear irreversible response: the contact area follows a complex decrease-increase-decrease path during unloading rather than retracing the loading branch, and repeatable petal-like edges appear, confirming a wrinkle-induced surface instability. Our results show that the observed contact area evolution is consistent with a competition between an area-increasing contribution and an area-reduction contribution. The latter is enhanced primarily by large deformation and dominates in softer PDMS, thereby masking the initial area increase and leading to a net reduction of contact area. The torsional contact results shown in the present work provide supporting experimental evidence that large deformation can drive shear-induced contact area reduction, as suggested by previous numerical studies \cite{lengiewicz2020finite, mergel2021contact}. However, the driving force for the initial area increase still remains unclear.

It should be noted that varying the PDMS mixing ratio changes not only the elastic modulus but also interfacial adhesion and bulk viscoelasticity. As for adhesion, it is well established that PDMS with a lower crosslink density exhibits a higher work of adhesion due to increased chain mobility and the presence of free chain ends at the surface \cite{vorvolakos2003effects}. In adhesive contact mechanics, a higher work of adhesion tends to resist contact area reduction under shear, as it imposes an additional resistance to crack propagation at the contact edge. Consequently, if adhesion were the primary driver of the observed area reduction, the softer PDMS samples (e.g., 1:15 and 1:20), which possess stronger adhesion, would be expected to show less contact area reduction. This expected trend, however, is not observed in our measurements (Figs. \ref{fig:Fig_7} and \ref{fig:Fig_8}), although direct adhesion or surface-energy measurements were not performed in the present work. Past studies indicate that the PDMS samples with the mixing ratio between 1:10 and 1:20 tend to reach a stabilized deformation under creep within 20 sec \cite{Zhang2024PDMS_relaxation,Johnston2019PDMSModulus}. Therefore, the dwell time of $2$ min used in the present study is expected to be sufficient to ensure that the contact area is already close to stabilization and becomes relatively insensitive to further creep if the torsional loading rate is sufficiently low \cite{zhang2024non}.

Although large deformation can explain the dominant area-reduction branch, experimentally proving that it is the main origin of the initial area increase itself is nearly impossible. A plausible candidate mechanism for that initial increase is tribo-electrification \cite{lacks2019long}, as previously suggested by Zhang et al. \cite{zhang2024non}. In future work, numerical models that incorporate contact electrification, viscoelasticity and large deformation can be developed to isolate and quantify its contribution during incipient torsional loading individually \cite{silavnieks2025flexoelectricity}. Additionally, the role of tribo-electrification can be probed experimentally by applying an external electric field during the torsional contact test, which may suppress or enhance the initial area increase depending on the polarity of the applied field.  

\section{Conclusion}\label{sec:Conclusion}
In this work, we experimentally investigated how loading direction and large deformation affect the torsional contact response of soft PDMS spheres. The loading-unloading torsional tests show that the interface is clearly irreversible: while the contact area increases first and then decreases during loading, the unloading path follows a distinct decrease-increase-decrease evolution rather than retracing the loading branch. During unloading, repeatable petal-like edges appear in the contact region, indicating a wrinkle-induced surface instability intrinsic to the soft elastomer.

By varying the PDMS mixing ratio, we further show that the contact-area evolution results from a competition between an area-increasing contribution and an area-reduction contribution. As the material becomes softer and undergoes larger deformation, the area-reduction contribution is strengthened and eventually suppresses the initial area increase, producing a monotonic area decrease during loading for mixing ratios of $1:15$ and $1:20$. At the same time, softer PDMS exhibits lower shear strength, weaker torque oscillations, and improved repeatability. These results provide experimental evidence that large deformation can drive shear-induced contact area reduction, in agreement with previous numerical predictions, while the origin of the initial contact area increase remains unresolved.

Overall, the present results narrow the possible mechanisms responsible for the initial area increase: large deformation, normal force, angular velocity, and creep are unlikely to be its primary cause. This points to other interfacial effects, such as tribo-electrification, as promising candidates for future study. More broadly, the torsional contact data reported here provide a stringent benchmark for frictional contact models, which must capture both the irreversibility of the loading-unloading cycle and the material-dependent, non-monotonic evolution of the contact area under shear.

\section*{Acknowledgments}
Authors thank Mr. Bo Zhang, Dr. Yue Ding, Dr. Julien Scheibert, Dr. Davy Delmas, Dr. Huixing Wei, and Dr. Weiting Chen for their inspiring discussions. YX thanks Prof. Ying Hu (Hefei University of Technology) for providing facilities to fabricate PDMS samples. This work was supported by the National Natural Science Foundation of China (No. 52105179) and the Fundamental Research Funds for the Central Universities (JZ2025HGTG0298). 

\begin{spacing}{1}
	\bibliographystyle{asmejour}
	\bibliography{ref}

@article{savkoor1977effect,
  title={The effect of tangential force on the contact of elastic solids in adhesion},
  author={Savkoor, AR and Briggs, GAD0364},
  journal={Proceedings of the Royal Society of London. A. Mathematical and Physical Sciences},
  volume={356},
  number={1684},
  pages={103--114},
  year={1977},
  publisher={The Royal Society London}
}

@article{waters2010mode,
  title={Mode-mixity-dependent adhesive contact of a sphere on a plane surface},
  author={Waters, Julie F and Guduru, Pradeep R},
  journal={Proceedings of the Royal Society A: Mathematical, Physical and Engineering Sciences},
  volume={466},
  number={2117},
  pages={1303--1325},
  year={2010},
  publisher={The Royal Society Publishing}
}

@article{mergel2018continuum,
  title={Continuum contact models for coupled adhesion and friction},
  author={Mergel, Janine C and Sahli, Riad and Scheibert, Julien and Sauer, Roger A},
  journal={The Journal of Adhesion},
  year={2019},
  volume={95},
  pages={1101-1133},
  publisher={Taylor \& Francis}
}

@article{sahli2018evolution,
  title={Evolution of real contact area under shear and the value of static friction of soft materials},
  author={Sahli, R and Pallares, G and Ducottet, C and Ben Ali, IE and Al Akhrass, S and Guibert, Matthieu and Scheibert, Julien},
  journal={Proceedings of the National Academy of Sciences of the USA},
  volume={115},
  number={3},
  pages={471--476},
  year={2018},
  publisher={National Acad Sciences}
}

@article{vorvolakos2003effects,
  title={The effects of molecular weight and temperature on the kinetic friction of silicone rubbers},
  author={Vorvolakos, Katherine and Chaudhury, Manoj K},
  journal={Langmuir},
  volume={19},
  number={17},
  pages={6778--6787},
  year={2003},
  publisher={ACS Publications}
}

@article{papangelo2019mixed,
  title={On mixed-mode fracture mechanics models for contact area reduction under shear load in soft materials},
  author={Papangelo, Antonio and Ciavarella, Michele},
  journal={Journal of the Mechanics and Physics of Solids},
  volume={124},
  pages={159--171},
  year={2019},
  publisher={Elsevier}
}

@article{mcmeeking2020interaction,
  title={The interaction of frictional slip and adhesion for a stiff sphere on a compliant substrate},
  author={McMeeking, Robert Maxwell and Ciavarella, Michele and Cricr{\`\i}, Gabriele and Kim, K S},
  journal={Journal of Applied Mechanics},
  volume={87},
  number={3},
  pages={031016},
  year={2020},
  publisher={American Society of Mechanical Engineers}
}

@article{xu2022asperity,
  title={An asperity-based statistical model for the adhesive friction of elastic nominally flat rough contact interfaces},
  author={Xu, Yang and Scheibert, Julien and Gadegaard, Nikolaj and Mulvihill, Daniel M},
  journal={Journal of the Mechanics and Physics of Solids},
  volume={164},
  pages={104878},
  year={2022},
  publisher={Elsevier}
}

@article{lengiewicz2020finite,
  title={Finite deformations govern the anisotropic shear-induced area reduction of soft elastic contacts},
  author={Lengiewicz, Jakub and de Souza, Mariana and Lahmar, Mohamed A and Courbon, C{\'e}dric and Dalmas, Davy and Stupkiewicz, Stanislaw and Scheibert, Julien},
  journal={Journal of the Mechanics and Physics of Solids},
  volume={143},
  pages={104056},
  year={2020},
  publisher={Elsevier}
}

@article{aymard2024designing,
  title={Designing metainterfaces with specified friction laws},
  author={Aymard, Antoine and Delplanque, Emilie and Dalmas, Davy and Scheibert, Julien},
  journal={Science},
  volume={383},
  number={6679},
  pages={200--204},
  year={2024},
  publisher={American Association for the Advancement of Science}
}

@article{mergel2021contact,
  title={Contact with coupled adhesion and friction: Computational framework, applications, and new insights},
  author={Mergel, Janine C and Scheibert, Julien and Sauer, Roger A},
  journal={Journal of the Mechanics and Physics of Solids},
  volume={146},
  pages={104194},
  year={2021},
  publisher={Elsevier}
}

@article{wei2024does,
  title={Does static friction information predict the onset of sliding for soft material?},
  author={Wei, Huixin and Wang, Zhiyong and Tu, Xinhao and Cheng, Xuanshi and Li, Linan and Wang, Shibin and Li, Chuanwei},
  journal={International Journal of Solids and Structures},
  volume={305},
  pages={113087},
  year={2024},
  publisher={Elsevier}
}

@article{wei2025characterizing,
  title={Characterizing Friction Coefficients of Soft Materials via Stick-Slip Data in Static Friction: Mechanism Analysis and Experimental Validation},
  author={Wei, Huixin and Fang, Wei and Wang, Shibin and Wang, Zhiyong and Lin, Zehui and Liao, Baopeng},
  journal={Mechanics of Materials},
  pages={105545},
  year={2025},
  publisher={Elsevier}
}

@article{zhang2024non,
  title={Non-monotonic evolution of contact area in soft contacts during incipient torsional loading},
  author={Zhang, Bo and de Souza, Mariana and Mulvihill, Daniel M and Dalmas, Davy and Scheibert, Julien and Xu, Yang},
  journal={Tribology Letters},
  volume={72},
  number={4},
  pages={132},
  year={2024},
  publisher={Springer}
}

@article{ahmad2024effect,
  title={Effect of Finger Orientation on Contact Stiffness and Area During Sliding},
  author={Ahmad, Jahangier and AliAbbasi, Easa and Sormoli, MReza Alipour and Basdogan, Cagatay},
  journal={IEEE Transactions on Haptics},
  year={2024},
  volume={18},
  number={1},
  pages={175-187},
  publisher={IEEE}
}

@article{segalman2005new,
  title={New approximations for elastic spheres under an oscillating torsional couple},
  author={Segalman, Daniel J and Starr, Michael J and Heinstein, Martin W},
  journal={Journal of applied mechanics},
  volume={72},
  number={5},
  pages={705--710},
  year={2005},
  publisher={American Society of Mechanical Engineers Digital Collection}
}

@article{deresiewicz1954contact,
  title={Contact of elastic spheres under an oscillating torsional couple},
  author={Deresiewicz, H},
  year={1954},
  volume={21},
  number={1},
  pages={52--56},
  journal={ASME Journal of Applied Mechanics},
  publisher={American Society of Mechanical Engineers}
}

@article{lacks2019long,
  title={Long-standing and unresolved issues in triboelectric charging},
  author={Lacks, Daniel J and Shinbrot, Troy},
  journal={Nature Reviews Chemistry},
  volume={3},
  number={8},
  pages={465--476},
  year={2019},
  publisher={Nature Publishing Group UK London}
}

@article{silavnieks2025flexoelectricity,
  title={Flexoelectricity driven elastic contact-separation model for triboelectrification},
  author={Silavnieks, Ulvis and Jing, Qingshen and Gadegaard, Nikolaj and Mulvihill, Daniel M and Xu, Yang},
  journal={Friction},
  volume={14},
  number={3},
  year={2025},
  pages = {9441115},
  publisher={Tsinghua University Press}
}

@article{xu2025analytical,
  title={Analytical solutions to the shear-induced anisotropic area reduction in frictional elastomer contact},
  author={Xu, Mingzhu and Zhang, Mengru and Chen, Weiting and Zhao, Ya-Pu},
  journal={Science China Physics, Mechanics \& Astronomy},
  volume={68},
  number={8},
  pages={284611},
  year={2025},
  publisher={Springer}
}

@article{ceglie2025contact,
  title={Contact area shrinkage and increase in wavy frictional sliding contacts},
  author={Ceglie, Marco and Violano, Guido and Portaluri, Luigi and Algieri, Luciana and Afferrante, Luciano and Scaraggi, Michele and Menga, Nicola},
  journal={Journal of the Mechanics and Physics of Solids},
  pages={106389},
  year={2025},
  publisher={Elsevier}
}

@article{venkatadri2023torsion,
  title={Torsion-induced stick-slip phenomena in the delamination of soft adhesives},
  author={Venkatadri, Tara K and Henzel, Thomas and Cohen, Tal},
  journal={Soft Matter},
  volume={19},
  number={13},
  pages={2319--2329},
  year={2023},
  publisher={Royal Society of Chemistry}
}

@article{zeng2020experimental,
  title={Experimental Study on the Sliding of WJ-8 Small Resistance Fastener Composite Pad},
  author={Zeng, Zhiping and Wang, Di and Liu, Fushan and Shuaibu, Abdulmumin A and Lin, Zhihua},
  journal={Advances in Civil Engineering},
  volume={2020},
  number={1},
  pages={1918043},
  year={2020},
  publisher={Wiley Online Library}
}

@article{scholz1976asperity,
  title = {The role of asperity indentation and ploughing in rock friction --- I: Asperity creep and stick-slip},
  author = {Scholz, C. H. and Engelder, J. T.},
  journal = {International Journal of Rock Mechanics and Mining Sciences \& Geomechanics Abstracts},
  volume = {13},
  number = {5},
  pages = {149--154},
  year = {1976},
  doi = {10.1016/0148-9062(76)90819-6},
  publisher = {Elsevier}
}

@article{nikas2003elastohydrodynamics,
  title = {Elastohydrodynamics and Mechanics of Rectangular Elastomeric Seals for Reciprocating Piston Rods},
  author = {Nikas, George K.},
  journal = {Journal of Tribology},
  volume = {125},
  number = {1},
  pages = {60--69},
  year = {2003},
  doi = {10.1115/1.1506316},
  publisher = {ASME}
}

@article{zeka2026normal,
  title={Normal contact of metainterfaces: the roles of finite size and microcontact interactions},
  author={Zeka, Donald and Blal, Nawfal and Fekak, Fatima-Ezzahra and Duval, Arnaud and Gravouil, Anthony and Scheibert, Julien},
  journal={Journal of the Mechanics and Physics of Solids},
  pages={106646},
  year={2026},
  publisher={Elsevier}
}

@article{Johnston2019PDMSModulus,
  author  = {Johnston, Ian D. and McCluskey, Darryl K. and Tan, Ching-Kong and Tracey, Mary C.},
  title   = {Mechanical characterization of bulk Sylgard 184 for microfluidics and microengineering},
  journal = {Journal of Micromechanics and Microengineering},
  volume  = {24},
  number  = {3},
  pages   = {035017},
  year    = {2014},
  doi     = {10.1088/0960-1317/24/3/035017}
}

@article{Lee2016EffectOfCuringAgentConcentrationPDMS,
  author  = {Lee, W. S. and Yeo, K. S. and Andriyana, A. and Shee, Y. G. and Mahamd Adikan, F. R.},
  title   = {Effect of cyclic compression and curing agent concentration on the stabilization of mechanical properties of PDMS elastomer},
  journal = {Polymer Testing},
  volume  = {96},
  pages   = {347--354},
  year    = {2016},
  doi     = {10.1016/j.polymertesting.2016.04.016}
}

@article{Cho2021PdmsYoungModulus,
  author  = {Cho, Han Saem and Moon, Hen-Young and Lee, Heung Soon and Kim, Young Tae and Jeoung, Shin-Chang},
  title   = {Formulation Prediction for Young's Modulus of Poly(dimethylsiloxane) by Spectroscopic Methods},
  journal = {Bulletin of the Korean Chemical Society},
  volume  = {42},
  number  = {9},
  pages   = {1225--1231},
  year    = {2021},
  doi     = {10.1002/bkcs.12352}
}

@article{Zhang2024PDMS_relaxation,
  title={Temporal evolution of mechanical properties in PDMS: A comparative study of elastic modulus and relaxation time for storage in air and aqueous environment},
  author={Zhang, Yuanmin and Adam, Casey and Rehnstrom, Henrik and Contera, Sonia},
  journal={Journal of the Mechanical Behavior of Biomedical Materials},
  volume={160},
  pages={106779},
  year={2024},
  doi={10.1016/j.jmbbm.2024.106779}
}
\end{spacing}

\end{document}